\documentclass[conference]{IEEEtran}
\usepackage{caption}
\usepackage{booktabs}
\usepackage{multirow}
\usepackage{float}
\usepackage[most]{tcolorbox}
\IEEEoverridecommandlockouts

\usepackage{cite}
\usepackage{amsmath,amssymb,amsfonts}
\usepackage{algorithmic}
\usepackage{graphicx}
\usepackage{textcomp}
\usepackage{xcolor}
\usepackage{caption}
\usepackage{hyperref}
\def\BibTeX{{\rm B\kern-.05em{\sc i\kern-.025em b}\kern-.08em
    T\kern-.1667em\lower.7ex\hbox{E}\kern-.125emX}}
\begin{document}

\newcommand{\sys}[0]{\textsc{Dynamite}}

\title{\sys{}: Real-Time Debriefing Slide Authoring through AI-Enhanced Multimodal Interaction}

\author{
\IEEEauthorblockN{
Panayu Keelawat\textsuperscript{\dag},
David Barron\textsuperscript{\dag},
Kaushik Narasimhan\textsuperscript{\dag},
Daniel Manesh\textsuperscript{\dag},\\
Xiaohang Tang\textsuperscript{\dag},
Xi Chen\textsuperscript{\ddag},
Sang Won Lee\textsuperscript{\dag},
Yan Chen\textsuperscript{\dag}
}

\vspace{1ex}

\IEEEauthorblockA{
\textsuperscript{\dag}\textit{Virginia Tech}, USA\\
\{panayu, dbarron410, kaushikn06, danielmanesh, xiaohangtang, sangwonlee, ych\}@vt.edu
}

\IEEEauthorblockA{
\textsuperscript{\ddag}\textit{University of Virginia}, USA\\
nbj7kt@virginia.edu
}
}

\maketitle
\IEEEpeerreviewmaketitle

\begin{abstract}
Facilitating class-wide debriefings after small-group discussions is a common strategy in ethics education. Instructor interviews revealed that effective debriefings should highlight frequently discussed themes and surface underrepresented viewpoints, making accurate representations of insight occurrence essential. Yet authoring presentations in real time is cognitively overwhelming due to the volume of data and tight time constraints. We present \sys{}, an AI-assisted system that enables semantic updates to instructor-authored slides during live classroom discussions. These updates are powered by semantic data binding, which links slide content to evolving discussion data, and semantic suggestions, which offer revision options aligned with pedagogical goals. In a within-subject in-lab study with 12 participants, \sys{} outperformed a text-based AI baseline in content accuracy and quality. Participants used voice and sketch input to quickly organize semantic blocks, then applied suggestions to accelerate refinement as data stabilized.

\end{abstract}


\section{Introduction}

\begin{figure}[t]
    \centering
    \includegraphics[width=\linewidth, height=10cm, keepaspectratio]{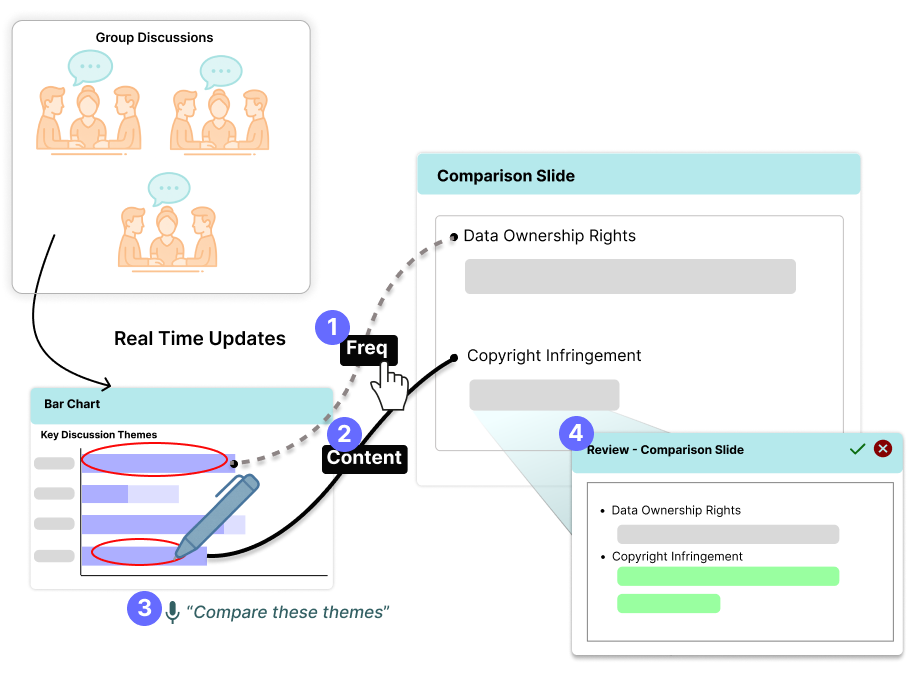}
    \caption{In a live Technology Ethics class, \sys{} enables semantic updates to slides based on evolving student-group discussion data. (1) Frequency-based and (2) content-based bindings link analytics data to semantic blocks on a slide. (3) Multimodal inputs like voice and sketching are used to generate and refine content. (4) As data evolves, goal-aligned semantic change suggestions appear for revision.}
    \label{fig:teaser}
\end{figure}

Instructors face significant challenges in managing live student group discussions, particularly in large virtual classrooms. This challenge is especially relevant in Technology Ethics courses, where debriefing is critical to help students synthesize diverse perspectives and engage in reflective reasoning. Technology Ethics courses, now widely integrated into Computer Science (CS) curricula \cite{b1,b2}, emphasize critical thinking and structured reasoning about the societal implications of technology \cite{b3,b4}. Unlike technical courses that emphasize problem-solving, ethics discussions require students to evaluate multiple perspectives, weigh ethical considerations, and construct well-reasoned arguments \cite{b5,b6}. Peer discussions in small groups are commonly adopted to expose students to diverse viewpoints and overlooked stakeholders, deepening their understanding of ethical dilemmas \cite{b7,b8,b9}. However, insights generated during these isolated group discussions often remain fragmented, posing a challenge when instructors attempt to synthesize and share them with the entire class. While instructors commonly conduct a class-wide debrief after small-group discussions, the process is often ad hoc due to the complexity and time constraints involved \cite{b10}. To understand what instructors look for in debriefings \cite{b6,b11}, we conducted interviews with six instructors. The interviews revealed that instructors care not only about the content of insights, but also about their frequency of occurrence. They want to systematically integrate diverse perspectives, connect student contributions to learning objectives, and facilitate deeper reflection. Even though AI-powered tools exist for generating slides \cite{b12}, crafting presentation materials is an iterative process that requires instructors to refine and adjust content as their understanding evolves. However, given the limited preparation time available (typically only 10 minutes per discussion session), instructors would face substantial cognitive demands if they attempted to rapidly examine live discussions, interpret evolving student insights, and select and refine content for presentation without technological support \cite{b13}. In large virtual classrooms, performing this synthesis effectively under such constraints would be nearly infeasible. To our knowledge, no existing tool accommodates the evolving, real-time nature of classroom discussions.

We conducted a preliminary study using an early version of the system with ten participants to better understand how people manage live discussion data given real-time analytics. Results showed that the resulting slides often contained errors due to outdated content, leading to irrelevant or misleading text. We also observed that participants struggled to keep up with the pace of the discussion. They had to interpret analytics, decide what to include, operate on live data, and continuously assess whether previous content remained relevant. To explore ways of supporting more effective iteration, we followed up with a paper prototype study involving six additional participants, focusing on alternative input modalities beyond traditional mouse clicks that might help users prepare slides more effectively.

We therefore propose \sys{}, an AI-assisted system designed to help instructors author debriefing slides during live student-group discussions (Fig. \ref{fig:teaser}). \sys{} has the ability to support semantic updates to slide content in response to evolving classroom insights. Semantic updates are enabled through two key mechanisms: semantic data binding, which links \textit{semantic blocks} on a slide to dynamic discussion data for automatic updates, and semantic suggestions, which provide context-aware revision options aligned with pedagogical goals. As new insights emerge, \sys{} evaluates whether the changes affect key learning dimensions derived from instructor interviews, and visualizes updates using slide-level badges. When a high-priority change is detected, the system notifies the instructor via a non-disruptive toast alert. To support the setup and refinement of semantic blocks, \sys{} offers multimodal input via voice and sketching. These inputs allow instructors to bind chart elements to slide regions and make quick adjustments to layout or content. A free-form authoring canvas displays live analytics as interactive charts, and semantic connections remain visible throughout, enabling users to validate or revise AI-generated content.

To evaluate \sys{}, we conducted a two-part within-subject in-lab study with 12 instructors. In the first part, participants used both \sys{} and a text-based AI baseline to complete slide authoring tasks based on predefined slide specifications. In the second part, participants engaged in an open-ended slide authoring task using \sys{}, followed by a semi-structured interview to reflect on their experience. Results showed that participants were able to quickly set up and validate semantic data bindings using multimodal input and visual connections for traceability. As a result, slides created with \sys{} were significantly more accurate than those created with the baseline, with a topic selection accuracy of 97.5\% compared to 75\%, and a layout accuracy of 91.6\% compared to 61.6\%. Once the structure was in place, participants used the remaining time to iteratively refine their slides, resulting in a significantly higher number of revisions (14.58 vs. 4.33 times). Importantly, participants were able to refine slides in later discussion sections, even if the data was not stabilized, because semantic suggestions offered low-effort, high-relevance edits that could be quickly applied.

To summarize, our work contributes:
\begin{enumerate}
    \item \textbf{\sys{}}, an AI-assisted system that enables semantic updates to debriefing slides in a virtual Ethics classroom.
    \item \textbf{Semantic updates}, integrating automatic updates through semantic data binding and instructor-guided revision through semantic suggestions.
    \item \textbf{Findings from an evaluation of \sys{}} through a within-subject user study (N=12), providing evidence for the system's effectiveness.

\end{enumerate}

By introducing \sys{}, we aim to provide an intelligent, scalable solution for managing live discussions and enhancing ethics education at scale. Our findings offer insights into how AI can support dynamic, group-based discussions without requiring the data to be finalized in a static form before analysis and synthesis.

\section{Related Work}

Our research builds upon prior work in three key areas, outlined below.

\subsection{Technology Ethics Education}

The integration of ethics into computing curricula underscores the critical role of structured reasoning and ethical reflection in the training of computer scientists~\cite{b1,b2,b4}. Pedagogical frameworks typically involve structured ethical reasoning exercises, such as analyzing ethical dilemmas through discipline-specific lenses~\cite{b3,b6}. To implement these activities, instructors often follow a multi-phase workflow: preparing relevant case scenarios or prompts before class, facilitating small-group or peer discussions during class, and leading a debriefing session where students collectively reflect on key ideas \cite{b4, b9}. However, despite widespread adoption, instructors often face challenges in synthesizing diverse student insights and facilitating comprehensive class-wide reflection. These difficulties stem from limited formal training, cognitive overload, and the inherent ambiguity in ethical discourse~\cite{b11,b6}.  

Various efforts have aimed to make ethics education more engaging and relatable through technology. For example, ethical reasoning has been blended into coding assignments to connect decision-making with technical work~\cite{b7,b14}, web-based simulations have been used to explore ethical trade-offs in AI decision-making~\cite{b8}, and interactive harm envisioning tools have been designed to highlight the real-world impact of AI systems~\cite{b15}. However, existing approaches primarily focus on enhancing student engagement, with little emphasis on technological support for instructors in real-time synthesis and debriefing of student-group discussions.

Unlike traditional ethics teaching approaches, \sys{} leverages multimodal AI-assisted interactions to facilitate the rapid integration of dynamic classroom discussions into structured, inclusive debriefing presentations.

\subsection{Support Tools for Student Group Discussions}

Real-time classroom collaboration tools have been developed to help instructors manage student discussions, monitor participation, and foster inclusive class-wide engagement \cite{newb1, newb2}. Prior research, such as \textit{VizGroup}~\cite{b16}, \textit{Groupnamics}~\cite{b17}, and \textit{ClassInSight}~\cite{b18}, has primarily focused on visualizing quantitative or objective classroom interactions, often in structured contexts such as programming courses or group-based project activities \cite{b19, newb3}. Similarly, \textit{MeetingVis} provides visual summaries of synchronous discussions, allowing retrospective analysis of meeting content to help users recall key discussion points~\cite{b20}. While research suggests that debriefing after collaborative learning activities enhances learning outcomes \cite{b21}, technological support for structured debriefing remains limited. For instance, \textit{PyramidApp} offers debriefing support, but its functionality is constrained to flagging student responses \cite{b13}.

Beyond visualization-based tools, chatbot-driven systems have been introduced to facilitate discussions. For instance, \textit{DebateBot} supports structured deliberative discussions on ethical topics but operates primarily within predefined conversational frameworks and moderation rules~\cite{b22}. \textit{EduChatbot} employs LLMs to foster collaborative learning by dynamically engaging with students~\cite{b23}. Similarly, Chiang et al. leverage LLMs to create a “devil’s advocate" chatbot that introduces diverse opinions into group discussions to challenge students’ viewpoints~\cite{b24}. 

Moreover, systems like \textit{DesignQuizzer} and \textit{Sketchy} explore user-driven refinement of design suggestions based on live input from the crowd~\cite{b25,b26}. The concept of “living material" is also relevant as it emphasizes the continuous revision of content as new information emerges. However, such systems typically focus on asynchronous contexts, such as narrative literature reviews or journalism~\cite{b27,b28}.

In contrast to existing tools, \sys{} directly facilitates the synthesis of live ethical discussions, where content evolves continuously, perspectives remain diverse, and definitive correct answers rarely exist. The multimodal interactions integrated into \sys{} uniquely enable educators to iteratively refine and synthesize ambiguous insights into coherent, dynamic debriefing presentations, supporting deeper reflection and engagement in real-time classroom settings.

\subsection{AI-Driven Interaction with Educational Workflows}

The integration of artificial intelligence into multimodal interaction has expanded users' ability to generate and iteratively refine content, often in real-time scenarios. Recent work, such as Yen et al.'s \textit{Code Shaping}, highlights how free-form, AI-interpreted sketches enable iterative refinement of complex content such as programming code~\cite{b29}. Similarly, \textit{GistVis} demonstrates AI-generated word-scale visualizations that reduce cognitive load during the interpretation of dense textual data, facilitating faster sensemaking~\cite{b30}. \textit{DesignPrompt} uses modalities such as images, colors, semantics and texts to fine-tune prompts for design exploration \cite{b31}. Researchers have also explored intelligent substrates for video-based collaboration, demonstrating how real-time multimodal inputs enhance interaction and content co-creation among remote collaborators~\cite{b32}.  

Further studies have examined the complementary relationship between educators and AI~\cite{b33}. Olshefski et al. have collected discussion transcripts from high school English classes to foster AI advancement~\cite{b34}. LLMs have also exhibited strong performance in ethical reasoning when provided with structured ethical frameworks~\cite{b35}. However, these explorations have also surfaced student concerns regarding AI-teacher collaboration~\cite{b36}, particularly in areas such as contextual understanding and potential biases in AI-generated responses~\cite{b37}. 

Our work explicitly integrates multimodal interaction and AI-driven content generation into real-time educational workflows. \sys{} ensures that multimodal interactions align with educators' pedagogical objectives, addressing student concerns by maintaining human oversight and providing contextual interpretation of dynamic classroom insights. Additionally, \sys{} preserves student privacy \cite{b38} by performing group-level data analytics with anonymized student identifiers.

By uniquely integrating insights from related areas, our work extends existing pedagogical and technological frameworks. The result is an innovative interactive approach to authoring dynamic debriefing materials that systematically elevates student-generated insights into structured class-wide discussions in real-time, addressing significant pedagogical and interaction design gaps identified in prior work.

\section{Instructor Interviews}

We conducted semi-structured interviews with six experienced instructors and teaching assistants at our institution. The participants included three Technology Ethics instructors with 15, 30, and 41 years of teaching experience, and three teaching assistants with 0.5, 1, and 3 years of experience, respectively. Each interview lasted approximately 30 minutes and focused on current debriefing practices, challenges in synthesizing student discussions, and priorities when designing debriefing slide decks.

Our analysis revealed several key findings:

\subsubsection{Implicit Reliance on Quantitative Insights}

Instructors often described trying to identify recurring or distinctive ideas across groups to structure class-wide discussion. Several emphasized the value of surfacing frequently mentioned topics to reflect shared concerns, or contrasting ideas to promote deeper thinking. Others described highlighting rare but insightful comments, using them as catalysts for broader reflection. These practices reflect an implicit reliance on quantitative factors like insight frequency, yet are typically performed in an ad-hoc and unsystematic way.

\subsubsection{Limited Visibility into Discussions}
All participants reported that they could not monitor every group in real time. They had to “hop between” breakout rooms, often missing key moments or important context. Two participants used lightweight tools such as Google Sheets to collect student responses, but even then, responses were uneven, and instructors found it difficult to review all input during class.

\subsubsection{Metrics for Effective Debriefing}

Participants consistently emphasized specific properties for an effective debriefing slide deck:

\begin{itemize}
    \item \textbf{Critical Thinking (CT)}: Encouraging students to think deeper and more critically about the topic \textit{(5/6 participants)}.
    \item \textbf{Opinion Diversity (OD)}: Highlighting a range of perspectives to show the ethical complexity of issues \textit{(6/6 participants)}.
    \item \textbf{Class Engagement (CE)}: Supporting class-wide discussions that encourage participation \textit{(4/6 participants)}.
    \item \textbf{CS Relevance (CR)}: Connecting discussion points to technical or disciplinary concepts in computer science \textit{(5/6 participants)}.
    \item \textbf{Ethics Relevance (ER)}: Grounding conversation in meaningful ethical concepts \textit{(3/6 participants)}.
\end{itemize}

\section{Preliminary Study}

\begin{figure}[b]
    \centering
    \includegraphics[width=\linewidth, height=10cm, keepaspectratio]{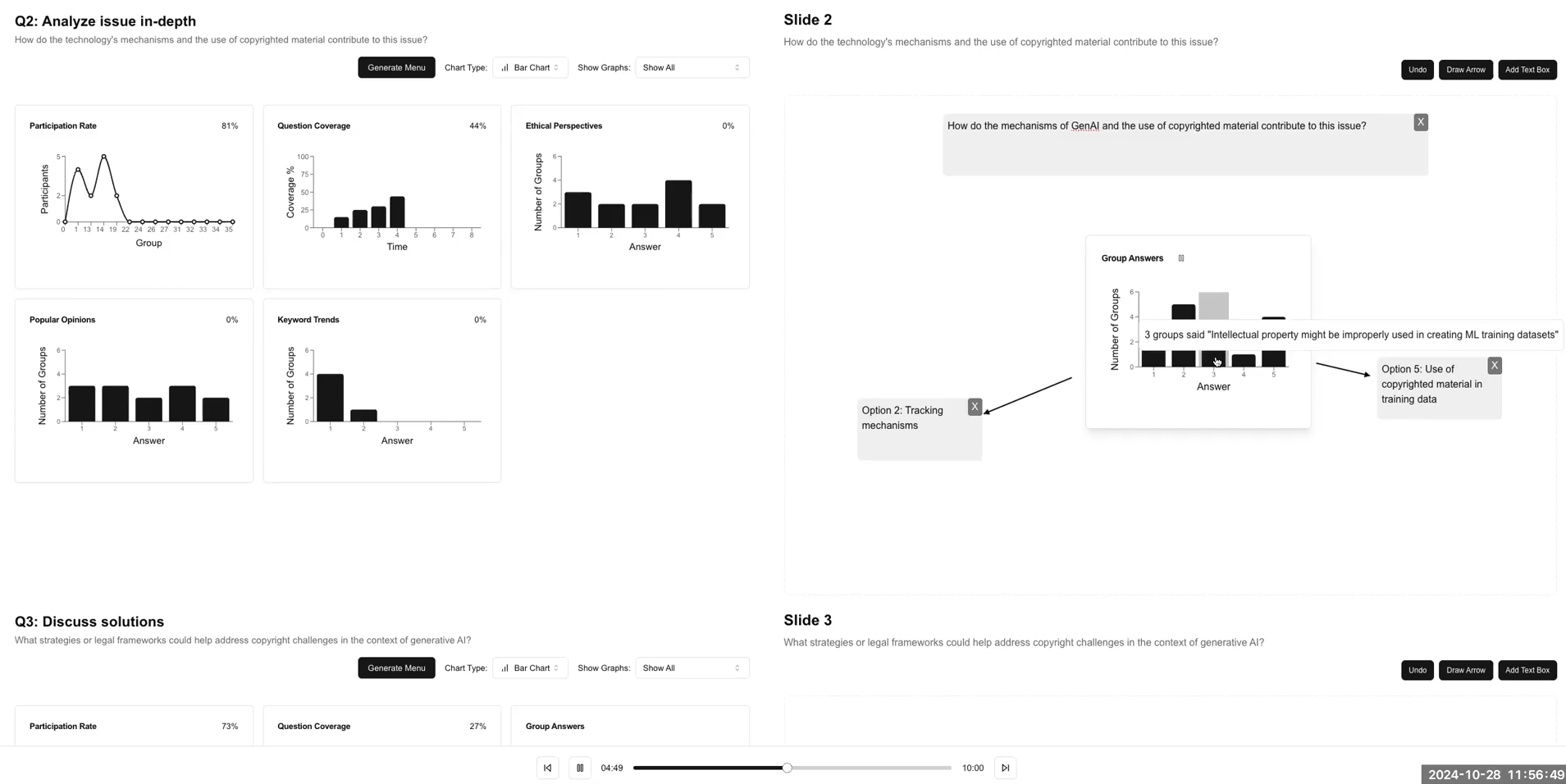}
    \caption{Screenshot of the early version of the system used in the preliminary study.}
    \label{fig:oldEthicQuest}
\end{figure}

We developed and evaluated an early version of the system with ten participants who had prior teaching experience. This prototype supported real-time chart-based analytics and used an LLM (GPT-4) to generate textual content from chart data with mouse clicking and typing (Fig.~\ref{fig:oldEthicQuest}).

We underestimated the time constraint instructors face in these settings. Typical group discussions lasted approximately 10 minutes, and most participants \textit{(6/10 participants)} barely completed the task, leaving no time for refinement. The remaining participants required an additional 2.20 minutes on average to finish their slides. We observed that participants first examined the data, then manually dragged relevant charts onto the slide before composing commentaries or talking points. When updates occurred, they often regenerated the text to reflect the changes, but this sometimes led to errors, particularly as participants rushed to prepare remaining slides. Due to the time pressure, participants relied heavily on live-updating analytics with minimal structure or explanation, resulting in slide decks that required improvisation during the debrief.

To address these constraints, we explored alternative input modalities that might accelerate the slide authoring process. We conducted a paper prototyping study to understand how instructors might intuitively express their intentions using different input modes. We recruited six additional participants with teaching experience and asked them to interact with a printed version of the \sys{} interface. Participants were instructed to describe how they would direct the system to generate and refine slides based on evolving discussion content, assuming the system could perfectly understand any of their input.

This study revealed how instructors naturally combine multimodal interactions, such as voice commands and sketch-based gestures, to express their authoring intent. These findings directly informed the design of \sys{}’s supported multimodal features, which we describe in a subsequent section.

\section{Design Goals}

Our instructor interviews and preliminary study revealed several key challenges driving our design goals. While instructors aim to help students develop ethical reasoning skills, they often struggle to construct structured debriefing materials due to limited access to discussion data from all groups, particularly in large virtual classrooms. Additionally, they have minimal time to prepare presentation materials, as debriefing must occur immediately after small-group discussions conclude. To address these challenges, we developed four design goals to guide the iterative development of \sys{}.

\begin{itemize}
    \item \textbf{DG1. Structuring debrief content to foster learning.}  
    Effective debriefing slides should capture a wide range of student perspectives and ethical reasoning strategies. \sys{} should support instructors in surfacing diverse viewpoints, organizing key discussion themes, and presenting insights in a way that promotes critical reflection and helps students learn from their peers.

    \item \textbf{DG2. Enabling rapid iteration on presentation content.}  
    Because discussion insights emerge continuously, \sys{} must allow instructors to efficiently capture, organize, and apply new data. The system should minimize manual operations and facilitate quick revisions, enabling instructors to focus on generating and refining content rather than performing repetitive tasks.

    \item \textbf{DG3. Managing evolving discussions and prioritizing key updates.}  
    As group discussions evolve, instructors need mechanisms to track and respond to changes without becoming overwhelmed. \sys{} should provide intelligent update management by prioritizing revisions that require attention while automatically handling those that do not.

    \item \textbf{DG4. Maintaining traceability between data and slide content.}  
    Instructors must trust that their slides accurately represent student contributions. \sys{} should make the relationship between slide content and its underlying discussion data visible and navigable, enabling verification, refinement, and confident reuse.
\end{itemize}

\begin{figure*}
    \centering
    \includegraphics[width=1\linewidth, height=12cm, keepaspectratio]{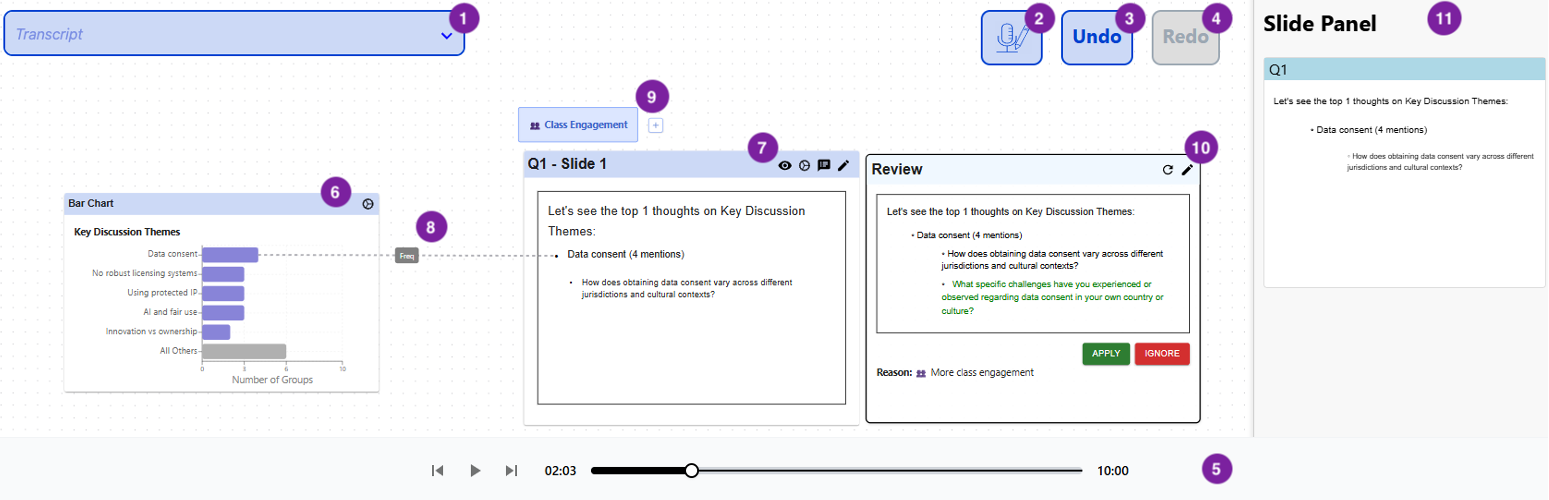}
    \caption{\sys{}'s interface. At the top of the workspace, a (1) transcription panel displays live speech, alongside controls for (2) multimodal input, (3) Undo, and (4) Redo. The bottom panel supports (5) playback of recorded classroom discussions and their associated analytics. The main canvas enables instructors to author slides, with (6) analytics nodes populating data in real time. (7) Slide nodes can be generated and bound to these visualizations with (8) semantic data binding. When there is a change suggestion to review, a (9) Suggested Change button appears, allowing instructors to review, accept, or modify updates via the (10) Review Panel. Slides are also listed in a right-hand (11) Slide Panel, where their order can be adjusted with drag-and-drop.}
    \label{fig:interface}
\end{figure*}

\section{\sys{} System}

\subsection{Live Classroom Analytics and Authoring Workspace}

\sys{} provides instructors with a unified, real-time interface that integrates classroom analytics and dynamic slide authoring within a single workspace (Fig. \ref{fig:interface}). At its core is a free-form canvas that supports flexible authoring of slide content during live discussions.
Analytics are organized by discussion points (e.g., Q1, Q2, Q3). For each point, \sys{} displays live-updating visualizations, such as bar charts and word clouds, that reflect emerging trends in student responses. By default, charts summarize data across three dimensions: Key Discussion Themes, Ethical Concepts, and Tech Topics (DG1). Additional charts, including Ethical Frameworks Used and Dimensions of Analysis, are available but hidden by default due to their more specialized use. All charts update continuously as new group responses arrive. Instructors can also gauge the overall discussion flow, as analytics activity typically begins with the first question and gradually shifts toward later discussion points (DG3). The canvas serves as a multimodal interaction surface. Instructors can issue commands via voice, sketching, or a combination of both \cite{b39}. Once slides are generated or refined based on these inputs, instructors can organize slide nodes and their connections by repositioning them freely on the canvas.

\begin{table*}[h]
\centering
\renewcommand{\arraystretch}{1.3}
\caption{Supported multimodal interactions in \sys{}.}
\label{table:interactions}
\begin{tabular}{p{3cm}|p{4.5cm}|p{7.5cm}}
\toprule
\textbf{Modalities} & \textbf{Interaction Type} & \textbf{Example Input or Gesture} \\
\midrule
\multicolumn{3}{c}{\textbf{Slide Generation}} \\
\midrule
\multirow{2}{*}{Voice} 
    & Frequency-based queries 
    & \begin{tabular}[t]{@{}l@{}}• \textit{“Generate a slide for the top three Key Discussion Themes”} \\• \textit{“Generate the least discussed Tech Topic”} \\• \textit{“Compare the highest and lowest frequencies of technology keywords} \\ \textit{mentioned"}\end{tabular} \\
\midrule
\multirow{3}{*}{Sketch + Voice} 
    & Generate slide from selected elements 
    & \begin{tabular}[t]{@{}l@{}}• Circle one chart element + \textit{“Talk about this idea”} \\• Circle two chart elements + \textit{“Compare these two keywords”} \end{tabular} \\
\cmidrule{2-3}
    & Retrieve relevant content from material 
    & \begin{tabular}[t]{@{}l@{}}• Circle a slide + \textit{“Find a relevant case study”}\end{tabular} \\
\midrule
\multicolumn{3}{c}{\textbf{Slide Refinement}} \\
\midrule
\multirow{3}{*}{Sketch} 
    & Refine layout structure 
    & \begin{tabular}[t]{@{}l@{}}• Draw vertical line on a slide to make content side-by-side \\ • Draw bullet points on a slide to display as bullet point structure \\• Draw circle on a slide to display live pie chart from slide content \end{tabular} \\
\midrule
Sketch + Voice 
    & Rewrite content
    & \begin{tabular}[t]{@{}l@{}}• Strikethrough over a content text + \textit{“Add more commentary”} \\• Strikethrough over a content text + \textit{“Make this a question”}\end{tabular} \\
\bottomrule
\end{tabular}
\end{table*}

\subsection{Multimodal Content Generation and Refinement}

The design of supported commands in \sys{} was informed by findings from our paper prototyping sessions as mentioned earlier. We prioritized observed interactions that are both fast and intuitive for instructors (DG2). Table~\ref{table:interactions} summarizes the supported modalities and example commands. 

\subsubsection{Slide Generation}

Voice-only commands are the fastest method for generating slides and were the most frequently mentioned in our paper prototype study. Instructors can issue high-level commands such as \textit{“Generate a slide for the top three Key Discussion Themes,”} \textit{“Generate the least discussed Tech Topic,”} or \textit{“Compare the highest and lowest frequencies of technology keywords mentioned.”} These commands typically reference frequency or category rank but omit details such as layout preference or content type (e.g., commentary versus class question). In such cases, \sys{} uses default values, typical bullet-point layouts and commentaries.

Combined voice-sketch commands allow instructors to specify more targeted operations. For example, instructors may circle two chart elements while saying, \textit{“Compare these two keywords,”} or brush on a single element and say, \textit{“Talk about this category.”} Unlike voice-only input, these interactions make visual references explicit, reducing ambiguity. In some cases, instructors may also circle an existing slide and issue a command like \textit{“Find a relevant case study,”} prompting the system to retrieve supporting material.

\subsubsection{Slide Refinement}

Layout refinement is another most popular operation found in the paper prototype study. It is performed using sketch input, especially immediately after generating a slide when the default layout does not match the instructor’s intent. Sketching allows for intuitive reorganization of content at a fine level of control. \sys{} supports three layout structures: bullet points, side-by-side comparison, and pie chart. These correspond to common instructional intents such as listing arguments, comparing viewpoints, or presenting class discussion analytics.

Content rewriting is another common refinement task. In this case, the instructor specifies both the type of content to modify (e.g., commentary or question) and the location on the slide. This is accomplished by brushing over a specific semantic content block while issuing commands such as \textit{“Add more commentary”} or \textit{“Make this a question.”}

\subsubsection{Multimodal Input Translator}

\sys{} translates multimodal input into actionable commands by analyzing voice and sketch gestures in context. For sketch input, the system identifies the element being brushed over and the sketch's spatial relation to that element. If the sketch matches a predefined pattern for layout refinement, the command is executed directly without invoking the language model. For voice input, the system parses utterances using predefined templates and extracts relevant parameters. Contextual references such as \textit{“this”} or \textit{“here”} are resolved by associating them with the sketched elements. Once all parameters are determined, they are passed to the \hyperref[sec:semantic-block-pipeline]{Semantic Block Generation Pipeline}, which synthesizes the requested update into a new semantic block to be rendered on the slide.

Instructors may also perform manual edits for unsupported operations, such as typing, deleting, or rearranging content directly on the slide without relying on multimodal input.

\subsection{Semantic Updates}

\sys{} supports semantic updates through semantic bindings and suggestion mechanisms that ensure slide content remains aligned with evolving discussion data and pedagogical goals. There are three core components:

\subsubsection{Semantic Blocks}

A semantic block is a unit of content generated from source analytics that expresses an idea as a qualitative statement, such as a class question or instructor commentary (DG1). Each semantic block is bound to a single source analytic and can be displayed in any slide region. Multiple semantic blocks can be generated from the same source analytic, allowing instructors to highlight different aspects of a single chart or insight.

\subsubsection{Semantic Binding}

Semantic binding links a slide region to a specific source analytic and determines how that connection behaves over time (DG4). \sys{} supports two types of semantic binding: content-based and frequency-based. Instructors can toggle between binding types by clicking the edge button attached to the connection.

\paragraph{Content-Based Binding}

This type of binding focuses on the content of the idea, regardless of its frequency in the chart. In other words, the binding is static in regards to time. Content-based binding is typically generated through multimodal interactions that directly reference specific content, such as sketching over a category and saying, \textit{“Talk about this category.”} In these cases, the user’s intent is tied to the idea itself, not its popularity.

\paragraph{Frequency-Based Binding}

Frequency-based binding links the slide region to a ranked position within a chart, e.g., the most frequent or least frequent category, rather than to a specific content label. This binding is dynamic: if the frequency ranking of categories changes, the bound source analytic and corresponding slide region update automatically to reflect the new data (DG3). For example, if a user requests, \textit{“Generate a slide about the most popular discussion theme,”} the system creates a frequency-based binding because the instruction prioritizes rank over specific content.

To ensure accuracy, \sys{} includes a \texttt{FrequencyMonitorModule} that tracks changes in occurrence rankings. When a bound insight’s position is displaced, the module updates the binding and triggers generation of semantic block in that slide region using the \hyperref[sec:semantic-block-pipeline]{Semantic Block Generation Pipeline}.

\subsubsection{Semantic Suggestions}

Semantic suggestions allow instructors to refine existing semantic blocks by surfacing overlooked or newly relevant information. This feature is designed to assist with ongoing slide refinement after initial setup, particularly as new data may shift the pedagogical relevance of slide content (DG3). Unlike frequency-based bindings, which update content automatically, suggestions require instructor approval before applying.

\paragraph{Slide Deck Evaluation}

Suggestions are ranked and prioritized using a component called \texttt{MetricEvaluator}, which analyzes both slide-level and deck-level metrics aligned with debriefing goals (DG1). Since \texttt{MetricEvaluator} has access to the Analytics Database, it can assess which records with their semantic attributes are already represented in the slide deck and identify areas for improvement.

\paragraph{Computing Suggestions}

The \texttt{SemanticSuggestionEngine} is invoked when the discussion activity for a given discussion point drops below 10\% of total group participation, or when less than two minutes remain in the session. These thresholds ensure that suggestions are computed at appropriate times when data is likely to have stabilized (DG3). With ranked metrics from \texttt{MetricEvaluator}, the engine identifies target slides, and then calls the \hyperref[sec:semantic-block-pipeline]{Semantic Block Generation Pipeline} to create suggested content updates.

\paragraph{Visualizing Change Suggestions}

When suggestions are available, a Suggested Change button appears above the target slide. If multiple suggestions are present, only the top-ranked one is shown; others can be accessed by clicking the adjacent \textit{“+”} button. When an instructor clicks the Suggested Change button, the system displays a Review Panel showing the content diff between the original and suggested semantic block. The instructor can then choose to apply, ignore, refresh, or modify the suggestion (DG2). If the Suggested Change button is not currently in view, a non-disruptive toast notification is triggered to nudge the instructor to review the update \cite{b40}.

\subsection{Technical Details}

\sys{} is implemented using ReactJS, React Flow, NodeJS, and Supabase. To enable semantic-level understanding and support content updates, the system relies on three core technical pipelines:

\subsubsection{Transcript Analysis Pipeline}

As real-time transcripts are streamed into the system, we apply topic modeling to new sentences with substantial length to extract relevant topics and subtopics for the current discussion point. Cosine similarity is then used to detect topic duplication. If a topic is identified as a duplicate, we update the cumulative time spent on that topic and associate any newly contributing groups with it in the database. If the topic is new, we classify it using discourse dimensions. Discourse classification is powered by GPT-4o and guided by dataset prompting using the DiscussionTracker corpus \cite{b34}. For each utterance, we store a set of semantic attributes in the database, including \texttt{isNewIdea}, \texttt{isAgreement}, \texttt{isChallenge}, \texttt{isEvidence}, and \texttt{isExtension}. We also record CS and Ethics keywords, if identified by predefined keyword set.

\subsubsection{Semantic Block Generation Pipeline}
\label{sec:semantic-block-pipeline}

\begin{figure}[t]
    \centering
    \includegraphics[width=\linewidth, height=10cm, keepaspectratio]{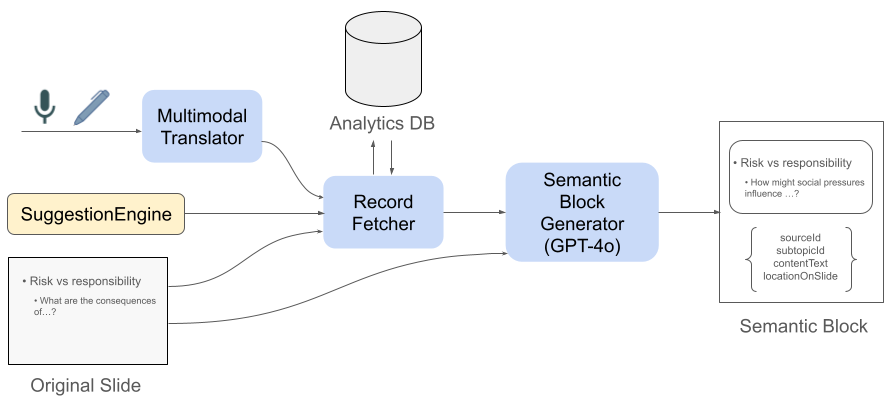}
    \caption{Semantic Block Generation Pipeline: Multimodal input or system triggers are parsed. The Record Fetcher queries the Analytics DB for relevant insights, which, along with the original slide, are passed to the Semantic Block Generator to produce a refined semantic block for slide integration.}
    \label{fig:semBlockPipeline}
\end{figure}

This pipeline can be invoked either through multimodal input or by the \texttt{SemanticSuggestionEngine} (Fig. \ref{fig:semBlockPipeline}). Once triggered, the system examines the specified debriefing metric to optimize and constructs a query to retrieve relevant records from the analytics database, scoped to the current discussion point and topic (if provided). If the metric is CT, the system prioritizes records labeled as \texttt{isEvidence} or \texttt{isChallenge}. If no such records are found, it selects the one with the highest time spent \cite{b41}. If the metric is OD, the pipeline first checks for the most frequently mentioned idea that is not already represented on the current slides. If that idea has already been included, it selects the least frequently mentioned yet unused idea. For CE, the system identifies records from groups that are least represented across the existing slide deck to ensure broader group coverage. If the metric is CR or ER, the system selects records mentioning the most frequently occurring domain-specific keyword. If the word has already been used, it selects the least-mentioned but still relevant keywords that are not yet present in the slide content. Finally, the selected records are passed to the \texttt{SemanticBlockGenerator}, a GPT-4o-powered module that synthesizes the content into a structured semantic block. These blocks are positioned within the slide region~\cite{b42}, based on both the retrieved data and the current slide context.

\subsubsection{Semantic Binding Update Pipeline}

To ensure a smooth user experience during updates, \sys{} takes a proactive approach by pre-generating potential semantic blocks and storing them in the browser's local storage. For each analytic chart, if a new category of insight emerges, the system uses the \hyperref[sec:semantic-block-pipeline]{Semantic Block Generation Pipeline} to prepare a corresponding block in advance. As the new insight's occurrence begins to approach the currently bound insight (i.e., a frequency difference of less than two), the system refreshes the pre-generated block with the latest data. When an actual update is triggered, \sys{} retrieves the corresponding semantic block from local storage and displays it seamlessly without glitching. This allows instructors to immediately review and respond to updates without interrupting their authoring workflow.

\section{Evaluation}

\subsection{Study Setting}

We conducted a two-part, in-lab study involving 12 participants (seven males, five females) with prior teaching experience and background in Ethics education. Participants were recruited via university mailing lists and personal networks.

To assess the system’s semantic updates and design, participants interacted with two prototype conditions:

\begin{itemize}
    \item \textbf{Baseline:} This condition featured a text-only interface for requesting slide content. When a participant submitted a text request, the system captured the current state of the visual charts and passed the query to the same large language model (GPT-4o) used in \sys{}. However, this version excluded multimodal interactions and the semantic update features.
    
    \item \textbf{\sys{}:} This condition offered full access to all multimodal and AI-assisted updates. To ensure familiarity, participants were provided with a video tutorial and a printed cheat sheet prior to beginning the tasks.
\end{itemize}

\subsubsection{Part 1: Structured Task Completion}

The first part of the study focused on comparing system performance and user perceptions across the Baseline and \sys{} conditions. Participants completed a series of debriefing slide authoring tasks using both systems in a within-subjects design.  For each condition, participants followed a predefined slide deck specification designed by the research team, which required creating five slides based on a ten-minute classroom discussion dataset. The slide specifications covered three discussion questions (Q1, Q2, Q3), with Q1 and Q2 each requiring two slides and Q3 requiring one slide. The order of conditions and datasets was counterbalanced to mitigate learning effects. Prior to each condition, participants were given 5-10 minutes to explore the system using a warm-up dataset. When ready, they initiated playback of the simulated live-streamed discussion and began preparing slides in real time. After the ten-minute authoring period, participants completed a NASA-TLX workload questionnaire. This procedure was repeated for the second condition.

\subsubsection{Part 2: Open-ended Exploration and Reflection}

The second part of the study aimed to gather qualitative insights in a more contextualized setting. Participants used \sys{} without a predefined slide specification, allowing for open-ended exploration. They were encouraged to explore features they had not used in Part 1 and reflect aloud on their choices and decision-making strategies during interaction. A think-aloud protocol was used throughout. Following this session, we conducted a semi-structured interview to elicit deeper insights and clarify observations.

Each study session lasted approximately 1.5 hours. Participants received a \$25 USD compensation. All sessions were screen- and audio-recorded for analysis.

\subsection{Results}

We conducted Wilcoxon signed-rank tests with Benjamini-Hochberg (BH) correction for multiple NASA-TLX metric comparisons (using Python’s SciPy library) between the Baseline and \sys{} conditions. Additionally, two researchers analyzed the study recordings. Results could be grouped into the following sections.

\subsubsection{Slide Content Analysis}

To evaluate how well the resulting slides met task expectations in the first part of the study, we assessed their completeness and accuracy against the predefined specifications. To illustrate how accuracy was determined, consider one task that required participants to create a slide comparing the most and least popular ethical frameworks side by side, with each topic accompanied by 1–2 bullet-point commentaries. Layout accuracy was determined by whether the slide used the correct side-by-side format and included the expected number of bullet points. Topic selection accuracy required that the topics reflected the highest and lowest frequency values at the 10-minute mark. Description accuracy was based on whether the bullet points included relevant commentaries. The results of this analysis are summarized in Table~\ref{tab:slide-accuracy}.

\begin{table}[t]
\centering
\caption{Slide Accuracy and Completeness Comparison}
\label{tab:slide-accuracy}
\begin{tabular}{@{}p{4cm}cc@{}}
\toprule
\textbf{Dimension} & \textbf{\sys{}} & \textbf{Baseline} \\
\midrule
Avg. Slides Generated (out of 5)     & \(5.00 \pm 0.00\)     & \(4.16 \pm 0.41\) \\
Layout Accuracy (\%)                & \(91.60 \pm 8.16\)    & \(61.60 \pm 9.83\) \\
Topic Selection Accuracy (\%)       & \(97.50 \pm 4.18\)    & \(75.00 \pm 8.94\) \\
Description Accuracy (\%)           & \(95.00 \pm 12.25\)   & \(70.00 \pm 4.92\) \\
\bottomrule
\end{tabular}
\vspace{1mm}
\end{table}

Slides created with \sys{} significantly outperformed those created with the baseline in both accuracy and completeness, based on Wilcoxon signed-rank tests (\textit{p} $<$ 0.05 for all metrics). Participants using \sys{} consistently generated the full set of five slides, while baseline participants produced fewer on average. \sys{} also yielded significantly higher accuracy overall. These results suggest that \sys{} was able to correctly surface the underlying semantic data bindings participants were looking for. The near-perfect topic selection accuracy indicates that participants were able to manage evolving data when guided by \sys{}. In contrast, slightly lower scores in layout and description accuracy likely reflect instances where participants overlooked specific formatting instructions or omitted suggested commentary and discussion questions. Nevertheless, \sys{} consistently led to higher performance across all metrics.

\subsubsection{Semantic Suggestions Provide Relevant Edits Fast}

\begin{figure}
    \centering
    \includegraphics[width=1\linewidth, height=5.25cm, keepaspectratio]{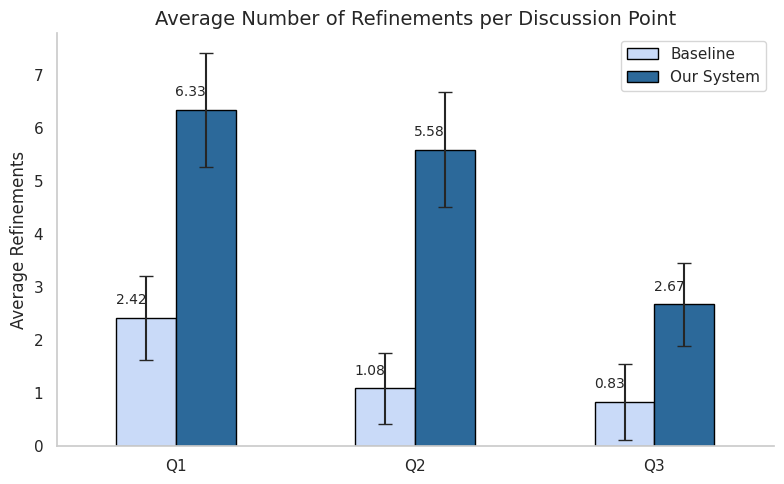}
    \caption{Average number of refinements per question under baseline and \sys{}. Participants made substantially more refinements when using our system across all three questions.}
    \label{fig:refinementComparison}
\end{figure}

Participants frequently used semantic suggestions during moments of idle time while waiting for new data to populate. Instead of just pausing, they revisited earlier slides once the underlying analytics for the question had stabilized, and applied semantic suggestions to improve clarity or completeness. This behavior was especially common among participants who had structured their slides early and were monitoring for final updates. As P2 explained, \textit{“While the data for question 2 is still coming in, I know it’s time to look at the question 1 slides. Then I can add more things because I know they’re final [for question 1].”} As a result, we observe in Fig.~\ref{fig:refinementComparison} that refinements are concentrated in Q1, with fewer adjustments made during Q2 and Q3. Although Q3 included only one slide by design, the trend is still apparent when comparing Q1 and Q2.

This return-and-refine pattern made semantic suggestions especially valuable for participants aiming to enhance detail without rewriting from scratch. Participants had quick access to meaningful change suggestions, and they reduced decision-making overhead, making it easier to finalize slides within tight time constraints. Several participants remarked that the suggestions surfaced ideas or wordings they hadn’t initially considered but ultimately appreciated. 

Interestingly, participants rarely focused on the suggestion labels such as “Critical Thinking" or “Ethics Relevance." Rather than using these categories to guide their choices, they evaluated the suggestion content directly based on whether the wording filled a perceived gap or improved clarity. Participants might also make small manual edits before or after accepting a suggestion, indicating that the tool was helpful even when the fit wasn’t perfect. Due to its speed and targeted application, the suggestion feature accounted for more than half of all refinement actions.


\subsubsection{Semantic Binding Enables Trust and Reduces Oversight}

A key advantage of \sys{} lies in its ability to maintain semantic bindings between slide content and the underlying data. Once participants verified that the bindings were correct, they expressed high confidence in the system’s ability to preserve accuracy during subsequent updates. This validation step served as a cognitive handoff: rather than continually monitoring system outputs, participants were able to delegate responsibility for maintaining alignment. While not statistically significant, perceived performance scores from the NASA-TLX were higher with \sys{} (mean = 2.33) compared to the baseline (mean = 3.58). One participant explained, \textit{“Basically, I just check it against the bar—if my ‘top three’ matches the longest bars, I move on.”} In contrast, the baseline condition required constant monitoring. Participants often had to regenerate entire slides when mismatches occurred, and several adopted a strategy of waiting for the underlying analytics to stabilize before prompting the system. These approaches were time-consuming and frequently led to incomplete slide sets with the baseline system.

Participants also used the semantic bindings as a tool to identify errors caused by the AI-enhanced system. In one specific example, when a user said \textit{“Generate a slide about the least discussed tech topic"} and the system misheard \textit{“topic"} as \textit{“top"}, the incorrect binding became visible in the connection between the chart and the slide. By tracing this connection, participants could identify that the system had linked the slide to the wrong bar in the chart. In response, they would repeat the voice command more clearly to correct the linkage. P4 recalled, \textit{“I think it got mixed up with other commands… it generated this [slide], and because I didn’t want it, I just tried again.”} This traceability made it easier for participants to detect and recover from mismatches without having to extensively read the content on the slide.

While semantic binding reduced the need for checking the correctness, we still observed participants keeping an eye on chart movement to decide when to begin refining slide content. This behavior highlights an opportunity to provide clearer indicators of when charts are considered stable or finalized.

\begin{figure}[t]
    \centering
    \includegraphics[width=1\linewidth]{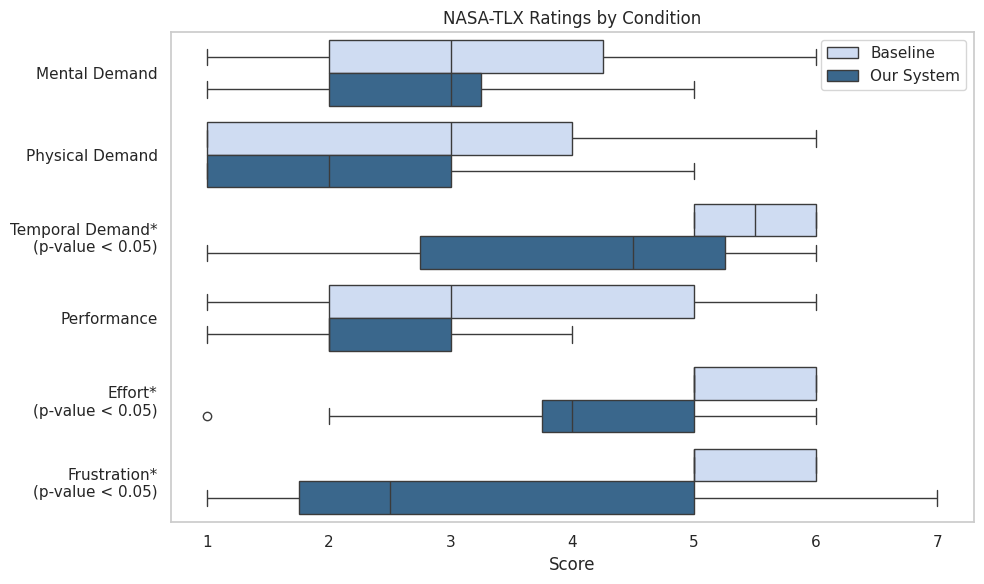}
    \caption{NASA-TLX results from Part 1 of the study. Participants reported significant lower level of temporal demand, effort and frustration.}
    \label{fig:nasatlx}
\end{figure}

Three metrics showed statistically significant differences: Temporal Demand, Effort, and Frustration (all corrected \textit{p} = 0.0468 $<$ 0.05). In other cases, \sys{} yielded lower mean scores than the baseline (Fig. \ref{fig:nasatlx}). These findings suggest that \sys{} meaningfully reduces oversight-related workload, enabling participants to keep pace with the discussion and feel less rushed during the task.

\subsubsection{Multimodal Input Enhances Speed and Flexibility}

The availability of voice and sketch-based input in \sys{} improved participants' ability to both structure and refine slides. Voice commands allowed users to rapidly generate entire slide frameworks, including semantic data bindings, in a single utterance, even if some details were omitted. P10 noted, \textit{“When I just say the voice command I don’t have to type. The slide shows up right away and saves me time.”} Sketching, in turn, facilitated precise, low-level adjustments such as selecting visual layouts and fine-tuning slide content. Participants also described sketching as intuitive and aligned with the layout metaphor; for instance, using a single circle to represent a pie chart.

Although some voice commands failed to register or produced errors, participants typically recovered quickly by reissuing the command. They could also use both voice and sketching at the same time to disambiguate their intent in several scenarios. Nevertheless, a common strategy involved using voice to scaffold initial slide content, then switching to sketching for refinement. 

\subsubsection{\textit{“I can imagine using this in the real classroom.”}}

All participants expressed interest in using \sys{} during actual classroom sessions, particularly for managing high-volume discussions in large online classes. They valued the system’s ability to structure content in real time and guide slide creation in ways that aligned with how they naturally organize ideas during a debrief. As P8 noted, \textit{“If there’s time, I’d be happy to look at the suggestions and keep editing my slides.”} Participants also suggested additional features, such as the ability to search for images and create custom charts. Overall, they were satisfied with the resulting slides and viewed \sys{} as a practical tool for surfacing student contributions and enhancing the quality of debriefings through more structured presentations.

\section{Discussion}

\subsection{Impact on Presentation Authoring Workflows with Live Data}

This study introduces a new authoring workflow that integrates dynamic classroom data, enabling instructors to create debriefing slides on the fly during live discussions. Instructors no longer need to wait for group work to conclude or manually review large volumes of student input. Instead, they can interact with structured insights as they emerge, allowing for more responsive and timely synthesis. This approach is especially valuable when working with qualitative data, where interpretation is often complex, subjective, and context-dependent compared to purely quantitative information.

\subsection{Applications Beyond the Technology Ethics Classroom}

The implications of \sys{} extend beyond Technology Ethics courses to any domain where open-ended, discussion-based learning is central. Courses in Art and Design, Humanities, and the Social Sciences also benefit from surfacing diverse viewpoints and fostering critical reflection. In these settings, a system that can capture, organize, and synthesize evolving student perspectives in real time could significantly enhance instructional practices. Additionally, the virtual classroom context of our study reveals design implications for broader deployment. Voice input was effective in online settings, but may feel awkward or disruptive in physical classrooms, highlighting the need to expand support for alternative modalities such as stylus input. Beyond education, the responsive authoring workflow demonstrated in \sys{} may also be applicable to domains such as live social media trend tracking, public deliberations, or town hall meetings, where synthesizing fast-changing qualitative input is essential.

\section{Limitations \& Future Work}

This study was conducted in an in-lab simulated classroom setting, which may not fully capture the pressure of a live classroom environment. Future work should explore deploying the system in actual classroom settings to better understand how time constraints, social dynamics, and instructional context influence instructors' use of the system.

While the system currently offers a predefined set of charts, future iterations could explore more customization features. Enabling instructors to define new visualizations or configure alerts for specific discussion topics could further personalize the experience. As noted in the Results section, some participants manually edited slides when AI-generated text did not match their intended tone or framing. This points to a future direction of adapting the system to mimic an instructor’s writing style, enabling even faster and more coherent iteration.

Finally, the use of real-time discussion monitoring raises important privacy considerations. Although \sys{} is designed to avoid displaying exact student transcripts, thereby minimizing the risk of spotlighting individuals, students may still self-censor if they perceive their conversations are being monitored. Future research should investigate how to balance transparency, privacy, and expressiveness in classroom-facing monitoring tools.

\section{Conclusion}

This work introduces \sys{}, an AI-assisted system designed to support instructors in authoring real-time debriefing slides based on dynamic, small-group discussions in virtual Technology Ethics classrooms. The system leverages semantic updates, powered by multimodal input, to manage evolving data throughout the session. Our within-subject evaluation with 12 participants demonstrates that \sys{} significantly improves slide accuracy and completeness compared to a baseline text-only AI system. Participants used voice and sketch commands to structure slides and bind content in real time, validate outputs by tracing the bindings, and hand off updates to the system. They revisited slides as data stabilized, efficiently refined content, and reviewed and applied change suggestions, enabling more rounds of iteration, even on the final slide in the deck. More broadly, \sys{} illustrates how AI-enhanced systems can shift educational workflows from static, post-hoc synthesis toward responsive, just-in-time content creation. Future work will explore how to scale this approach to more complex pedagogical settings and enhance the personalization and transparency of AI-generated content.

\section*{Acknowledgment}

We thank Prof. Dan Dunlap and our collaborators in the Computer Science and Philosophy departments for their valuable feedback and support throughout this project. We also thank Sam Wong for his assistance with figure design. This study was approved by our institution’s IRB.

\clearpage
\appendices
\section{Instructional Sessions Used in Evaluation}
\label{appendix:datasets}

\noindent
We used Session 1 and 2 for the first part of the study. The other session was used in the second part of the study.

\vspace{0.5em}
\tcbset{colback=gray!10, colframe=black, boxrule=0.5pt, arc=2pt, left=2pt, right=2pt, top=2pt, bottom=2pt}

\noindent
\begin{tcolorbox}[title=Session 1: Generative AI and Copyright]
\noindent\textbf{Duration:} 10 minutes

\noindent\textbf{Instructions:} Please discuss and answer each question.

\noindent\textbf{Scenario:} Generative AI systems, like those used to create art, music, and text, have raised significant concerns about copyright infringement. In this exercise, we will explore the relationship between generative AI and copyright issues. Students will discuss the underlying problem, the causes of these copyright challenges, and brainstorm possible solutions to navigate the ethical and legal complexities in this area.

\noindent\textbf{Discussion Points:}
\begin{enumerate}
  \item What are the primary causes of copyright infringement concerns with generative AI?
  \item How do the technology's mechanisms and the use of copyrighted material contribute to this issue?
  \item What strategies or legal frameworks could help address copyright challenges in the context of generative AI? (E.g., how can creators and AI developers ensure fair use and protect intellectual property rights?)
\end{enumerate}

\noindent\textbf{Task:} Each group member should contribute their ideas on the topic and participate in the discussion. Afterward, collectively summarize the key points and prepare one question for class discussion, such as  
\textit{“What changes should be made to copyright laws to address the unique challenges posed by generative AI?”}

\rule{\linewidth}{0.4pt}

\noindent\textbf{Course Level:} Undergraduate \\
\textbf{Number of Students:} 85 students (23 groups)
\end{tcolorbox}

\vspace{-0.5em}
\noindent
\begin{tcolorbox}[title=Session 2: Technology and Surveillance in China]
\noindent\textbf{Duration:} 10 minutes

\noindent\textbf{Instructions:} Please discuss and answer each question.

\noindent\textbf{Scenario:} China has been increasing its use of surveillance technologies. Discuss this article from The New York Times \textit{“\href{https://www.nytimes.com/2022/06/21/world/asia/china-surveillance-investigation.html?referringSource=articleShare}{Four Takeaways From a Times Investigation Into China’s Expanding Surveillance State}.”}

\noindent\textbf{Discussion Points:}
\begin{enumerate}
  \item How do governments balance security and freedom?
  \item What level of surveillance do you think is appropriate?
  \item What are the benefits and drawbacks of such an intrusive and extensive program?
\end{enumerate}

\noindent\textbf{Task:} Each group member should contribute ideas during the discussion. Afterwards, please summarize the key points for each discussion point collectively.

\rule{\linewidth}{0.4pt}

\noindent\textbf{Course Level:} Graduate \\
\textbf{Number of Students:} 35 students (9 groups)
\end{tcolorbox}

\vspace{-0.5em}
\noindent
\begin{tcolorbox}[title=Session 3: Laser Weapons and AI-Controlled Robotic Weapons]
\noindent\textbf{Duration:} 10 minutes

\noindent\textbf{Instructions:} Please discuss and answer each question.

\noindent\textbf{Scenario:} Emerging military technologies now include laser weapons and AI-powered robotic systems. Discuss the scoop from WIRED \textit{“\href{https://www.wired.com/story/us-military-robot-drone-guns/}{The AI Machine Gun of the Future Is Already Here}.”}

\noindent\textbf{Discussion Points:}
\begin{enumerate}
  \item Do laser weapons, given their precision, make war more ethical or more dangerous?
  \item Should robots be trusted to decide when to use deadly force in war?
  \item Should there be global rules or treaties for using robotic and laser weapons?
\end{enumerate}

\noindent\textbf{Task:} Each group member should contribute ideas during the discussion. Afterwards, please summarize the key points for each discussion point collectively.

\rule{\linewidth}{0.4pt}

\noindent\textbf{Course Level:} Graduate \\
\textbf{Number of Students:} 34 students (9 groups)
\end{tcolorbox}


\begin{thebibliography}{00}
\bibitem{b1} ABET, ``Criteria for Accrediting Computing Programs, 2024 – 2025,'' 2024. [Online]. Available: https://www.abet.org/accreditation/accreditation-criteria/criteria-for-accrediting-computing-programs-2024-2025

\bibitem{b2} T. Force, ``CS2023: ACM/IEEE-CS/AAAI Computer Science Curricula,'' 2024. [Online]. Available: https://csed.acm.org

\bibitem{b3} S. Tolmeijer, M. Kneer, C. Sarasua, M. Christen, and A. Bernstein, ``Implementations in machine ethics: A survey,'' \textit{ACM Comput. Surv.}, ACM, 2021, DOI:10.1145/3419633

\bibitem{b4} N. Brown, B. Xie, E. Sarder, C. Fiesler, and E. S. Wiese, ``Teaching ethics in computing: A systematic literature review of ACM computer science education publications,'' \textit{ACM Trans. Comput. Educ.}, ACM, 2024, DOI:10.1145/3634685

\bibitem{b5} R. Reich, M. Sahami, J. M. Weinstein, and H. Cohen, ``Teaching computer ethics: A deeply multidisciplinary approach,'' in \textit{Proc. SIGCSE 2020}, ACM, 2020, DOI:10.1145/3328778.3366951

\bibitem{b6} B. Moskal, K. Miller, and L. A. S. King, ``Grading essays in computer ethics: Rubrics considered helpful,'' in \textit{Proc. 33rd SIGCSE Tech. Symp. Comput. Sci. Educ.}, ACM, 2002, DOI:10.1145/563340.563380

\bibitem{b7} E. Peck, ``The ethical engine: Integrating ethical design into intro computer science,'' 2017. [Online]. Available: https://medium.com/bucknell-hci/the-ethical-engine-integratingethical-design-into-intro-to-computer-science-4f9874e756af

\bibitem{b8} E. Awad, S. Dsouza, R. Kim, J. F. Schulz, J. Henrich, A. F. Shariff, J. Bonnefon, and I. Rahwan, ``The Moral Machine experiment,'' \textit{Nature}, 2018, DOI:10.1038/s41586-018-0637-6

\bibitem{b9} B. J. Grosz, D. G. Grant, K. Vredenburgh, J. Behrends, L. Hu, A. Simmons, and J. Waldo, ``Embedded EthiCS: Integrating ethics across CS education,'' \textit{Commun. ACM}, ACM, 2019, DOI:10.1145/3330794

\bibitem{b10} P. Dillenbourg, ``Integrating technologies into educational ecosystems,'' \textit{Distance Educ.}, Taylor \& Francis, 2008.

\bibitem{b11} J. T. Chowning and P. Fraser, ``An ethics primer,'' Northwest Association of Biomedical Research, 2007.

\bibitem{b12} Gamma App, ``You have ideas. Gamma brings them to life. Powered by AI.,'' 2025. [Online]. Available: https://gamma.app

\bibitem{b13} M. Dieckmann, D. Hernández-Leo, and I. Amarasinghe, ``Flagging in teacher-facing orchestration dashboards: Factors affecting its use in pyramid CSCL debriefing,'' in \textit{Proc. ICALT 2022}, IEEE, 2022, DOI:10.1109/ICALT55010.2022.00047

\bibitem{b14} S. A. Doore, C. Fiesler, M. S. Kirkpatrick, E. Peck, and M. Sahami, ``Assignments that blend ethics and technology,'' in \textit{Proc. 51st ACM Tech. Symp. Comput. Sci. Educ.}, ACM, 2020, DOI:10.1145/3328778.3366994

\bibitem{b15} Z. J. Wang, C. Kulkarni, L. Wilcox, M. Terry, and M. Madaio, ``Farsight: Fostering responsible AI awareness during AI application prototyping,'' in \textit{Proc. CHI 2024}, ACM, 2024, DOI:10.1145/3613904.3642335

\bibitem{newb1} T. Wu, X. Tang, S. Wong, X. Chen, C. A. Shaffer, and Y. Chen, ``The impact of group discussion and formation on student performance: An experience report in a large CS1 course,'' in \textit{Proc. SIGCSE 2025}, ACM, 2025, DOI:10.1145/3641554.3701973

\bibitem{newb2} A. Y. Wang, Y. Chen, J. J. Y. Chung, C. Brooks, and S. Oney, ``PuzzleMe: Leveraging peer assessment for in-class programming exercises,'' \textit{Proc. ACM Hum.-Comput. Interact.}, vol. 5, no. CSCW2, Art. no. 415, pp. 1--24, Oct. 2021, DOI:10.1145/3479559

\bibitem{b16} X. Tang, S. Wong, K. Pu, X. Chen, Y. Yang, and Y. Chen, ``VizGroup: An AI-assisted event-driven system for collaborative programming learning analytics,'' in \textit{Proc. UIST 2024}, ACM, 2024, DOI:10.1145/3654777.3676347

\bibitem{b17} A. J. Sato, Z. Sramek, and K. Yatani, ``Groupnamics: Designing an interface for overviewing and managing parallel group discussions in an online classroom,'' in \textit{Proc. CHI 2023}, ACM, 2023, DOI:10.1145/3544548.3581322

\bibitem{b18} T. J. Ngoon, S. Sushil, A. E. B. Stewart, U.-S. Lee, S. Venkatraman, N. Thawani, P. Mitra, S. Clarke, J. Zimmerman, and A. Ogan, ``ClassInSight: Designing conversation support tools to visualize classroom discussion for personalized teacher professional development,'' in \textit{Proc. CHI 2024}, ACM, 2024, DOI:10.1145/3613904.3642487

\bibitem{b19} A. G. Zhang, Y. Chen, and S. Oney, ``VizProg: Identifying misunderstandings by visualizing students’ coding progress,'' in \textit{Proc. CHI 2023}, ACM, 2023, DOI:10.1145/3544548.3581516

\bibitem{newb3} X. Tang, X. Chen, S. Wong, and Y. Chen, ``VizPI: A real-time visualization tool for enhancing peer instruction in large-scale programming lectures,'' in \textit{Adjunct Proc. UIST 2023}, ACM, 2023, DOI:10.1145/3586182.3616632

\bibitem{b20} Y. Shi, C. Bryan, S. Bhamidipati, Y. Zhao, Y. Zhang, and K.-L. Ma, ``MeetingVis: Visual narratives to assist in recalling meeting context and content,'' \textit{IEEE Trans. Vis. Comput. Graph.}, 2018.

\bibitem{b21} I. Amarasinghe, D. Hernández-Leo, K. Manathunga, J. C. Pérez, and Y. Dimitriadis, ``Teacher-led debriefing in computer-supported collaborative learning pyramid scripts,'' in \textit{Proc. CSCL}, 2022, pp. 171--178.

\bibitem{b22} S. Kim, J. Eun, J. Seering, and J. Lee, ``Moderator chatbot for deliberative discussion: Effects of discussion structure and discussant facilitation,'' \textit{Proc. ACM Hum.-Comput. Interact.}, ACM, 2021, DOI:10.1145/3449161

\bibitem{b23} Z. Cai, S. Park, N. Nixon, and S. Doroudi, ``Advancing knowledge together: Integrating large language model-based conversational AI in small group collaborative learning,'' in \textit{Ext. Abstr. CHI 2024}, ACM, 2024, DOI:10.1145/3613905.3650868

\bibitem{b24} C.-W. Chiang, Z. Lu, Z. Li, and M. Yin, ``Enhancing AI-assisted group decision making through LLM-powered devil’s advocate,'' in \textit{Proc. IUI 2024}, ACM, 2024, DOI:10.1145/3640543.3645199

\bibitem{b25} Z. Peng, Q. Chen, Z. Shen, X. Ma, and A. Oulasvirta, ``DesignQuizzer: A community-powered conversational agent for learning visual design,'' \textit{Proc. ACM Hum.-Comput. Interact.}, vol. 8, no. CSCW1, Art. no. 44, pp. 1--40, Apr. 2024, DOI:10.1145/3637321

\bibitem{b26} S. Wallace, B. Le, L. A. Leiva, A. Haq, A. Kintisch, G. Bufrem, L. Chang, and J. Huang, ``Sketchy: Drawing inspiration from the crowd,'' \textit{Proc. ACM Hum.-Comput. Interact.}, ACM, 2020, DOI:10.1145/3415243

\bibitem{b27} R. Fok, A. Siu, and D. S. Weld, ``Toward living narrative reviews: An empirical study of the processes and challenges in updating survey articles in computing research,'' in \textit{Proc. CHI 2025}, ACM, 2025, DOI:10.1145/3706598.3714047

\bibitem{b28} M. Conlen and J. Heer, ``Idyll: A markup language for authoring and publishing interactive articles on the web,'' in \textit{Proc. UIST 2018}, ACM, 2018, DOI:10.1145/3242587.3242600

\bibitem{b29} R. Yen, J. Zhao, and D. Vogel, ``Code shaping: Iterative code editing with free-form AI-interpreted sketching,'' in \textit{Proc. CHI 2025}, ACM, 2025, DOI:10.1145/3706598.3713822

\bibitem{b30} R. Zou, Y. Tang, J. Chen, S. Lu, Y. Lu, Y. Yang, and C. Ye, ``GistVis: Automatic generation of word-scale visualizations from data-rich documents,'' in \textit{Proc. CHI 2025}, ACM, 2025, DOI:10.1145/3706598.3713881

\bibitem{b31} X. Peng, J. Koch, and W. E. Mackay, ``DesignPrompt: Using multimodal interaction for design exploration with generative AI,'' in \textit{Proc. DIS 2024}, ACM, 2024, DOI:10.1145/3643834.3661588

\bibitem{b32} H. Xia, T. Wang, A. Gunturu, P. Jiang, W. Duan, and X. Yao, ``CrossTalk: Intelligent substrates for language-oriented interaction in video-based communication and collaboration,'' in \textit{Proc. UIST 2023}, ACM, 2023, DOI:10.1145/3586183.3606773

\bibitem{b33} J. Jeon and S. Lee, ``Large language models in education: A focus on the complementary relationship between human teachers and ChatGPT,'' \textit{Educ. Inf. Technol.}, Springer, 2023.

\bibitem{b34} C. Olshefski, L. Lugini, R. Singh, D. Litman, and A. Godley, ``The Discussion Tracker Corpus of Collaborative Argumentation,'' in \textit{Proc. LREC 2020}, Marseille, France: European Language Resources Association, 2020, pp. 1033--1043. [Online]. Available: https://aclanthology.org/2020.lrec-1.130/

\bibitem{b35} A. S. Rao, A. Khandelwal, T. Kumar, U. Agarwal, and M. Choudhury, ``Ethical reasoning over moral alignment: A case and framework for in-context ethical policies in LLM,'' in \textit{Findings of the Assoc. Comput. Linguistics: EMNLP 2023}, ACL, 2023. [Online]. Available: https://aclanthology.org/2023.findings-emnlp.892

\bibitem{b36} B. Han, S. Coghlan, G. Buchanan, and D. McKay, ``Who is helping whom? Student concerns about AI-teacher collaboration in higher education classrooms,'' \textit{Proc. ACM Hum.-Comput. Interact.}, vol. 9, no. 2, May 2025, Art. no. CSCW206, pp. 1--32. DOI:10.1145/3711104

\bibitem{b37} U. Agarwal, T. Kumar, A. Khandelwal, and M. Choudhury, ``Ethical reasoning and moral value alignment of LLMs,'' in \textit{Proc. LREC-COLING 2024}, ELRA and ICCL, 2024. [Online]. Available: https://aclanthology.org/2024.lrec-main.560

\bibitem{b38} W. Kang, M. A. Hughes, and D. Roy, ``Anonymization of voices in spaces for civic dialogue: Measuring impact on empathy, trust, and feeling heard,'' \textit{Proc. ACM Hum.-Comput. Interact.}, vol. 8, no. CSCW2, Art. no. 482, pp. 1--22, Nov. 2024, DOI:10.1145/3687021

\bibitem{b39} S. G. Samuelsson and M. Book, ``Towards a visual language for sketched expression of software IDE commands,'' in \textit{Proc. VL/HCC 2023}, IEEE, 2023, DOI:10.1109/VL-HCC57772.2023.00021

\bibitem{b40} K. Pu, D. Lazaro, I. Arawjo, H. Xia, Z. Xiao, T. Grossman, and Y. Chen, ``Assistance or disruption? Exploring and evaluating the design and trade-offs of proactive AI programming support,'' in \textit{Proc. CHI 2025}, ACM, 2025, DOI:10.1145/3706598.3713357

\bibitem{b41} S. E. Toulmin, \textit{The Uses of Argument}. Cambridge, U.K.: Cambridge Univ. Press, 2003.

\bibitem{b42} R. Cheng, T. Barik, A. Leung, F. Hohman, and J. Nichols, ``BISCUIT: Scaffolding LLM-generated code with ephemeral UIs in computational notebooks,'' in \textit{Proc. VL/HCC 2024}, IEEE, 2024, DOI:10.1109/VL/HCC60511.2024.00012


\end{thebibliography}
\end{document}